\documentclass[a4paper,twoside,twocolumn,english,prl,intlimits,showpacs]{revtex4}
\usepackage{times}
\usepackage[T1]{fontenc}
\usepackage[latin2]{inputenc}
\usepackage{amsmath}
\usepackage{graphicx}
\usepackage{amssymb}

\makeatletter

\providecommand{\LyX}{L\kern-.1667em\lower.25em\hbox{Y}\kern-.125emX\@}

\usepackage{babel}
\makeatother
\begin{document}

\title{Lévy Flights in Inhomogeneous Media}

\author{D.~Brockmann}

\author{T.~Geisel}

\affiliation{Max-Planck-Institut für Strömungsforschung, Bunsenstraße 10, Göttingen,
Germany}

\begin{abstract}
We investigate the impact of external periodic potentials on superdiffusive
random walks known as Lévy flights and show that even strongly superdiffusive
transport is substantially affected by the external field. Unlike
ordinary random walks, Lévy flights are surprisingly sensitive to
the shape of the potential while their asymptotic behavior ceases
to depend on the Lévy index $\mu $. Our analysis is based on a novel
generalization of the Fokker-Planck equation suitable for systems
in thermal equilibrium. Thus, the results presented are applicable
to the large class of situations in which superdiffusion is caused
by topological complexity, such as diffusion on folded polymers and
scale-free networks.
\end{abstract}


\maketitle

\newcommand{\dy}{\mathrm{d}y}

\newcommand{\dx}{\mathrm{d}x}

\newcommand{\dk}{\mathrm{d}k}

\newcommand{\pdt}{\partial _{t}\, }

\newcommand{\fraclpmu}[1]{\Delta ^{\mu /2}}

\newcommand{\fou}[1]{\widetilde{#1}}

Diffusion processes are ubiquitous in nature. A freely diffusive particle
is characterized by a mean square displacement which increases linearly
in time, $\left\langle X^{2}(t)\right\rangle \propto t$. However,
a variety of interesting physical systems violate this temporal behavior.
For example, the position $X(t)$ of a superdiffusive particle heuristically
evolves as $X(t)\sim t^{1/\mu }$ with $0<\mu <2$. Superdiffusion
has been observed in a number of systems ranging from early discoveries
in intermittent chaotic systems~\cite{geise_00616:1985}, fluid particles
in fully developed turbulence~\cite{porta_01017:2001}, to millennial
climate changes~\cite{ditle_01441:1999} bacterial motion~\cite{levan_00237:1997}
and human eye movements~\cite{brock_xxxxx:xx}. 

Among the most successful theoretical concepts which have been applied
to superdiffusive phenomena is a class of random walks known as Lévy
flights~\cite{shles:1995,shles_00031:1993}. In constrast to ordinary
random walks, the displacements $\Delta x$ of a Lévy flight lack
a well defined variance, due to a heavy tail in the single step probability
density. Lévy flight models have paved the way towards a description
of superdiffusive phenomena in terms of fractional Fokker-Planck equations
(FFPE)~\cite{metze_00001:2000}. 

Since many of the aforementioned systems evolve in inhomogeneous environments,
it is crucial to understand the influence of external potentials on
the dynamics. While in ordinary diffusive systems an external force
is easily incorporated into the dynamics by a drift term in the corresponding
Fokker-Planck equation (FPE)~\cite{gardi:1985} the matter is more
subtle in superdiffusive systems due to the non-local properties of
the fractional operators involved. Depending on the underlying physical
model, different types of FFPEs are appropriate~\cite{brock_00409:2002},
therefore the ad hoc introduction of fractional operators may lead
to severe problems.

In cases where the external inhomogeneity can be represented by an
additive force, considerable progress has been made in a generalized
Langevin approach~\cite{jespe_02736:1999,foged_01690:1998} which
led to an FFPE in which deterministic and stochastic motion segregate
into independent components. This approach, however, is only suited
for systems in which such a segregation can be justified on physical
grounds. Furthermore, it is only valid for systems far from thermodynamic
equilibrium and fails in the large class of systems which obey Gibbs-Boltzmann
thermodynamics, e.g. where superdiffusion is caused by the complex
topology on which the process evolves, such as diffusion on folded
polymers~\cite{sokol_00857:1997}.

In this Letter we investigate the impact of external potentials on
this class of systems. Based on the paradigmatic case of a randomly
hopping particle on a folded copolymer, we report a number of bizarre
phenomena which emerge when Lévy flights evolve in periodic potentials
and show that external potentials have a profound effect on the superdiffusive
transport. This is in sharp contrast to generalized Langevin dynamics,
which displays trivial asymptotic behavior, a possible reason why
Lévy flights in periodic potentials have attracted little attention
in the past. We demonstrate that even strongly superdiffusive Lévy
flights are highly susceptible to periodic potentials. At low temperatures,
they exhibit a significant dependence on the overall shape of the
potential. Counterintuitively, the asymptotic behavior does not depend
on the Lévy flight index $\mu $, yet differs for various types of
potentials (except in the ordinary diffusion limit). A perturbation
analysis reveals a universal behavior for high temperatures. Finally
we show that in finite systems the effect of the potential on the
generalized diffusion coefficient is least pronounced for intermediate
values of $\mu $. This is consistent with the observation that Lévy
flights with $\mu \approx 1$ are particularly efficient in search
processes~\cite{brock_xxxxx:xx,viswa_00911:1999}. The results we
present are a first step towards an understanding of superdiffusive
dynamics on topologically complex structures exposed to external inhomogeneities. 

Let us begin with the dynamics of a particle performing an unbiased
random walk in a homogeneous environment in continuous time. The probability
$p(x,t)$ of finding the particle at a position $x$, given that it
was initially at the origin is governed by the master equation~\cite{gardi:1985},\begin{equation}
\pdt p(x,t)=\int \dy \left[w(x|y)\, p(y,t)-w(y|x)\, p(x,t)\right],\label{eq:master}\end{equation}
in which $w(x|y)$ is the probability rate of initiating a jump $y\rightarrow x$.
If this probability has a typical variance in distance one may expand
the rhs of Eq.~(\ref{eq:master}) in moments of $w(x|y)$ yielding
the FPE for a freely diffusive particle, $\pdt p=\Delta \, p$. However,
when the rate asymptotically follows an inverse power law of distance,
i.e $w(x|y)\sim |x-y|^{-(1+\mu )}$ with $\mu <2$ the variance of
jump lengths diverges and the particle performs a superdiffusive walk
known as a Lévy flight~\cite{shles:1995}. Inserting this rate into~(\ref{eq:master})
the rhs defines the integral operator $\fraclpmu{}$,\begin{equation}
\fraclpmu{}p(x,t)=\int \dy \frac{\left[p(y,t)-p(x,t)\right]}{|x-y|^{1+\mu }},\label{eq:fractionallaplace}\end{equation}
 and Eq.~(\ref{eq:master}) may be rewritten as\begin{equation}
\pdt p=D\fraclpmu{}p.\label{eq:freesup}\end{equation}
The parameter $D$ is the generalized diffusion coefficient. Up to
a constant factor, the operator $\fraclpmu{}$ is frequently refered
to as the fractional Laplacian, because it represents a multiplication
by $-|k|^{\mu }$ in Fourier space~\cite{metze_00001:2000}. This
simple spectral property of the operator is the reason why fractional
evolution equations are frequently introduced in the Fourier domain.
However, the position representation, Eq.~(\ref{eq:fractionallaplace}),
along with Eq.~(\ref{eq:master}), provides a more intuitive picture
of the dynamics and emphasizes the essential fact that $\fraclpmu{}$
is defined by a non-local integral kernel which decreases algebraically
with distance. Eq.~(\ref{eq:freesup}) is solved by $p(x,t)=(Dt)^{1/\mu }\, L_{\mu }\left(x/(Dt)^{1/\mu }\right)$
where $L_{\mu }(z)=(2\pi )^{-1}\int \dk \, \exp (ikz-|k|^{\mu })$
is the symmetric Lévy stable law of index $\mu $. The argument $x/t^{1/\mu }$
in $L_{\mu }$ reflects the superdiffusive behavior of the process.
When $\mu =2$ ordinary diffusion is recovered. In ordinary diffusion,
a potential $V$ is canonically introduced by a drift term $\beta \nabla V^{\prime }p$
in the FPE. Thus, it may seem reasonable to formally allow for an
external potential in a superdiffusive system by\begin{equation}
\pdt p=\beta \nabla V^{\prime }p+\fraclpmu{}p.\label{eq:frakfokker1}\end{equation}
This type of FFPE has been studied extensively in the past~\cite{jespe_02736:1999,foged_01690:1998}.
It describes deterministic motion in a gradient field $F=-\beta V^{\prime }$
subjected to Lévy stable white noise~$\eta (t)$, i.e. $\dot{X}=-\beta V^{\prime }+\eta (t)$.
However, this approach introduces severe restrictions. Systems evolving
according to~(\ref{eq:frakfokker1}) do not obey ordinary Gibbs-Boltzmann
thermodynamics. The stationary state $p_{s}$, if it exists, is generally
not $p_{s}\propto \exp (-\beta V)$ and depends on the tail parameter
$\mu $. Detailed balance is violated, and only in the diffusion limit
($\mu =2$) can the parameter $\beta $ be interpreted as an intensive
inverse temperature. 

The asymptotics of Eq.~(\ref{eq:frakfokker1}) in periodic potentials
is trivial. Rescaling the original coordinates $x,\, t\rightarrow z=x/\gamma ,\, \tau =t/\gamma ^{\mu }$
with $\gamma \gg 1$ yields a form invariant FFPE in a new potential
$\widehat{V}(z)=\gamma ^{\mu -2}\, V(\gamma \, z).$ The factor $\gamma ^{\mu -2}\ll 1$
implies that on large spatiotemporal scales any bounded potential
is insignificant to the dynamics, a possible explanation why Lévy
flights in periodic potentials have attracted little attention in
the past.

However, the generalized Langevin description is not appropriate for
a variety of superdiffusive phenomena~\cite{brock_00409:2002}, as
a segregation into deterministic and stochastic forces cannot be justified
by the underlying physics. %
\begin{figure}
\noindent \includegraphics[  width=1.0\columnwidth]{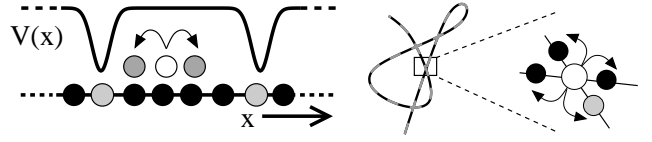}

\caption{\label{figcap:fig1}A particle jumps randomly along a folded copolymer
which consists of two types of periodically arranged monomers. Each
monomer type has a different potential sketched qualitatively in the
top left. Due to conformational changes of the chain the particle
can jump to a site close in Euclidean space yet distant in chemical
coordinate $x$.}
\end{figure}
 Consider the system depicted in Fig.~\ref{figcap:fig1}. A particle
is loosely attached to a polymer chain. Thermal activation causes
the particle to jump between monomers. The heterogeneity of the polymer
is accounted for by the potential $V(x)$ defined on chemical axis
$x$. It is reasonable to assume that the rate $w(x\pm a|x)$ of making
a transition between neighboring sites $x$ and $x\pm a$ decreases
with increasing potential at the target site and that it depends on
the potential difference $\delta V_{\pm }=V(x\pm a)-V(x)$ in units
of $k_{B}T=\beta ^{-1}$. Both assumptions are accounted for by: $w(x\pm a|x)\propto \exp [-\beta \delta V_{\pm }/2]$.
If the polymer is in solution and subjected to fast conformational
changes, regions of the chain that are distant along the chemical
axis $x$ of the polymer may come close in Euclidean space. Long distance
transitions $x\rightarrow y$ may occur with a probability which follows
an inverse power law of chemical distance $|x-y|^{-(1+\mu )}$ when
$|x-y|\gg a$. The exponent $\mu $ is determined by the folding properties
of the polymer, e.g. a Gaussian chain implies $\mu =1/2$~\cite{bouch_00127:1990}.
The possibility of initiating distant jumps along the chemical coordinate
enhances the diffusion process considerably, and is believed to play
a role in protein dynamics on DNA strands~\cite{berg_06929:1981}.
A jump process of this type can be modeled by a master equation~(\ref{eq:master}),
in which thermodynamic as well as geometric aspects need to be incorporated
in the rate $w(x|y)$. The above reasoning suggests,\begin{equation}
w(x|y)\propto e^{-\beta \left[V(x)-V(y)\right]/2}/|x-y|^{1+\mu }.\label{eq:inhomrate}\end{equation}
In Ref.~\cite{brock_yyyyy:yy}, a mean field theoretic treatment
of the dynamics of the polymer chain also yields Eq.~(\ref{eq:inhomrate}).
Inserting the above rate into Eq.~(\ref{eq:master}), we obtain\begin{equation}
\pdt p=e^{\beta V/2}\fraclpmu{}e^{-\beta V/2}\, p-p\, e^{\beta V/2}\fraclpmu{}e^{-\beta V/2},\label{eq:toposuper}\end{equation}
which is clearly different from the FFPE corresponding to generalized
Langevin dynamics. Eq.~(\ref{eq:toposuper}) obeys Gibbs-Boltzmann
thermodynamics, $p_{s}\propto \exp (-\beta V)$ is the stationary
solution, detailed balance is fulfilled, and $\beta $ is a well defined
intensive inverse temperature for all $\mu \in (0,2]$. Rescaling
coordinates as above yields a potential $\hat{V}(z)=V(\gamma z)$
lacking the pre-factor $\gamma ^{2-\mu }$ which is present in the
generalized Langevin scheme. Therefore, the effect of a bounded potential
will have an effect on all scales. Note that for $\mu =2$ Eq.~(\ref{eq:toposuper})
reduces to the ordinary FPE. When $V\equiv 0$ the rhs is identical
to $\fraclpmu{}p.$ 

Letting $\psi (x,t)=\exp [-\beta V(x)/2]\, p(x,t)$, Eq.~(\ref{eq:toposuper})
can be recast into a fractional Schrödinger equation,\begin{align}
\pdt \psi  & =-\mathcal{H}\, \psi \label{eq:ham1}\\
\mathcal{H} & =-\fraclpmu{}+U,\qquad U=e^{\beta V/2}\fraclpmu{}e^{-\beta V/2}\label{eq:ham2}
\end{align}
with an anomalous kinetic term $-\fraclpmu{}$ and an effective potential
$U$ which depends on $\mu $. A separation ansatz yields the associated
stationary equation,\begin{equation}
\left[E+\fraclpmu{}-U\right]\psi (x)=0\label{eq:statschroedinger}\end{equation}
for the spectrum $E$. Let us consider periodic potentials of wavelength
$2\pi \lambda $, $V(x)=V(x+2\pi \lambda n)$, with $n\in \mathbb{Z}$.
Without loss of generality we restrict ourselves to potentials with
vanishing offset and unit variance. A Bloch ansatz $\psi _{q}(x)=e^{iqx}\theta (x)$
with $\theta (x)=\theta (x+2\pi \lambda n)$ and $q\in [0,1/\lambda ]$
inserted into~(\ref{eq:ham1}) in Fourier space yields\begin{equation}
\left(E_{n,q}-E_{n,q}^{0}-\fou{U}_{0}\right)\fou{\theta }_{n}-\sum _{m\neq n}\fou{U}_{n-m}\fou{\theta }_{m}=0\label{eq:bands}\end{equation}
\begin{figure}
\noindent \includegraphics[  width=0.50\columnwidth]{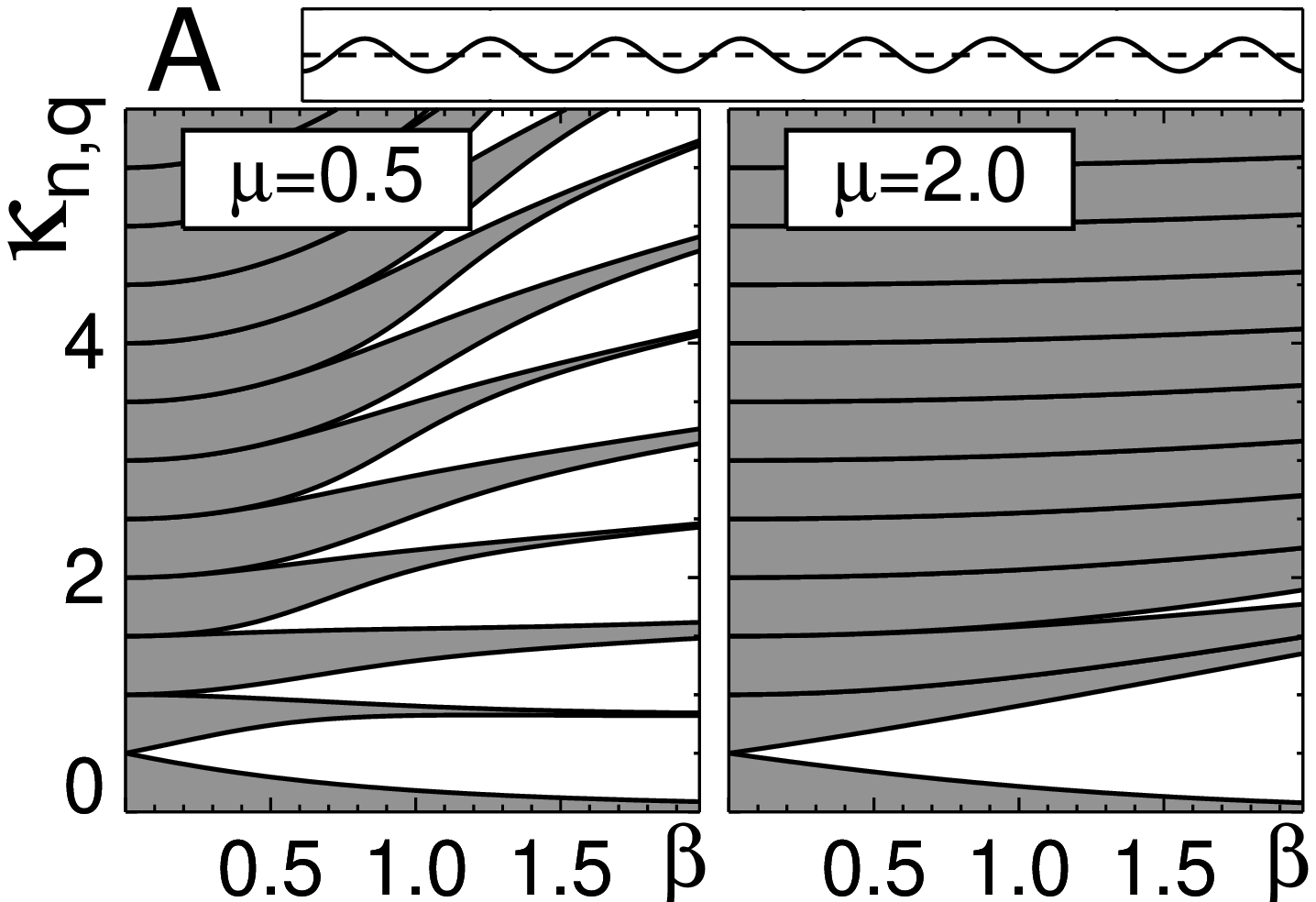}\hfill{}\includegraphics[  width=0.50\columnwidth]{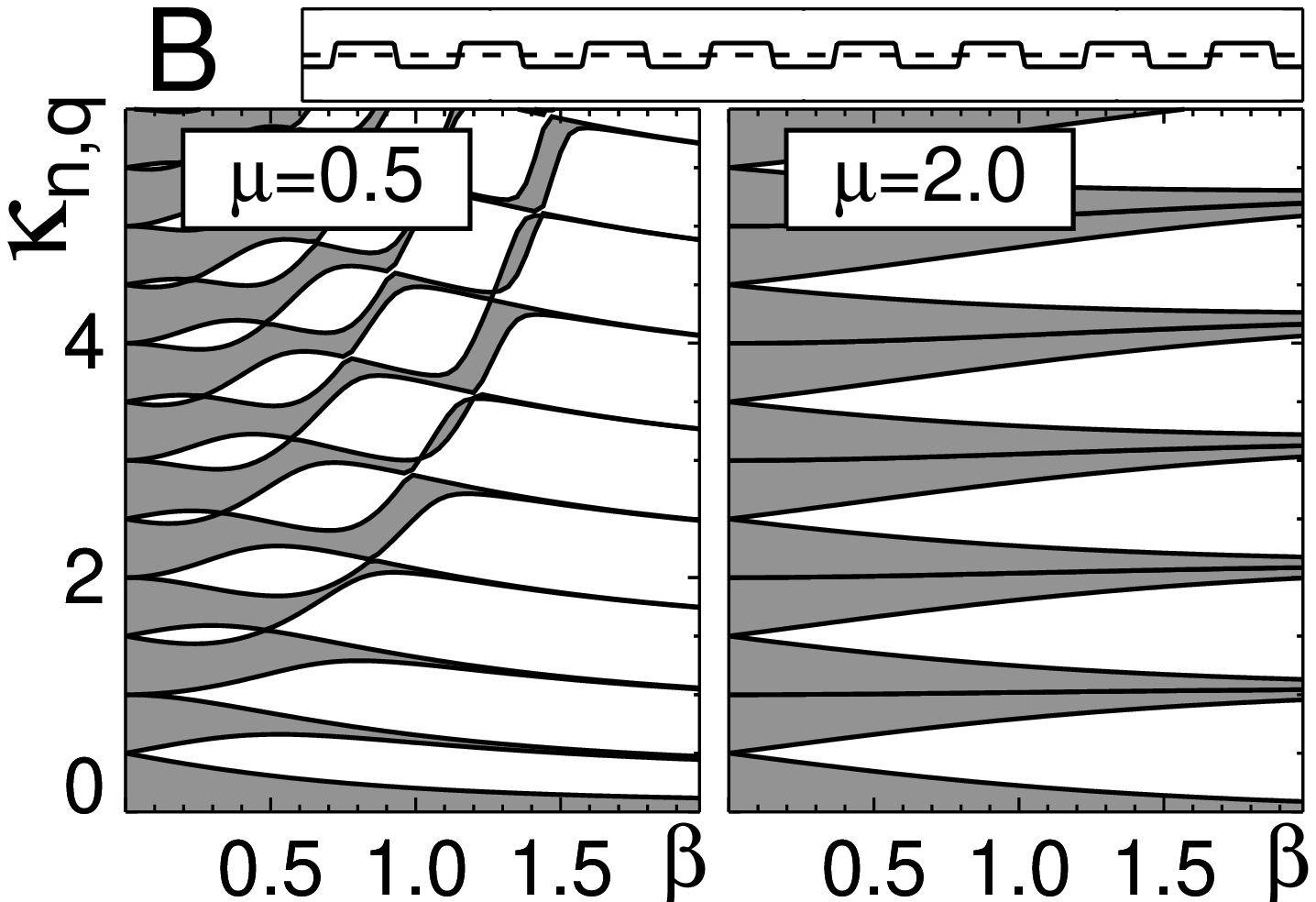}

\medskip{}
\noindent \includegraphics[  width=0.50\columnwidth]{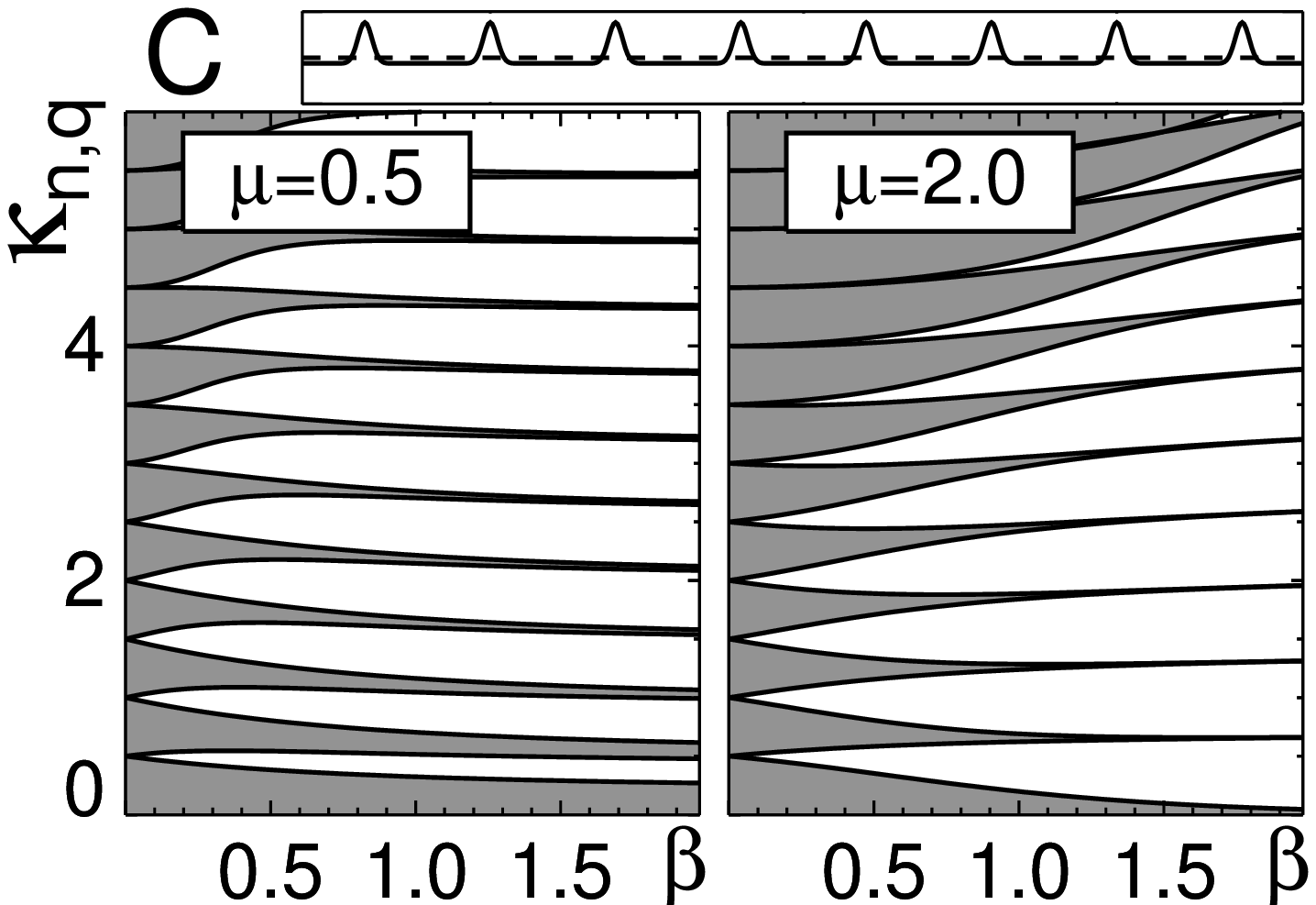}\hfill{}\includegraphics[  width=0.50\columnwidth]{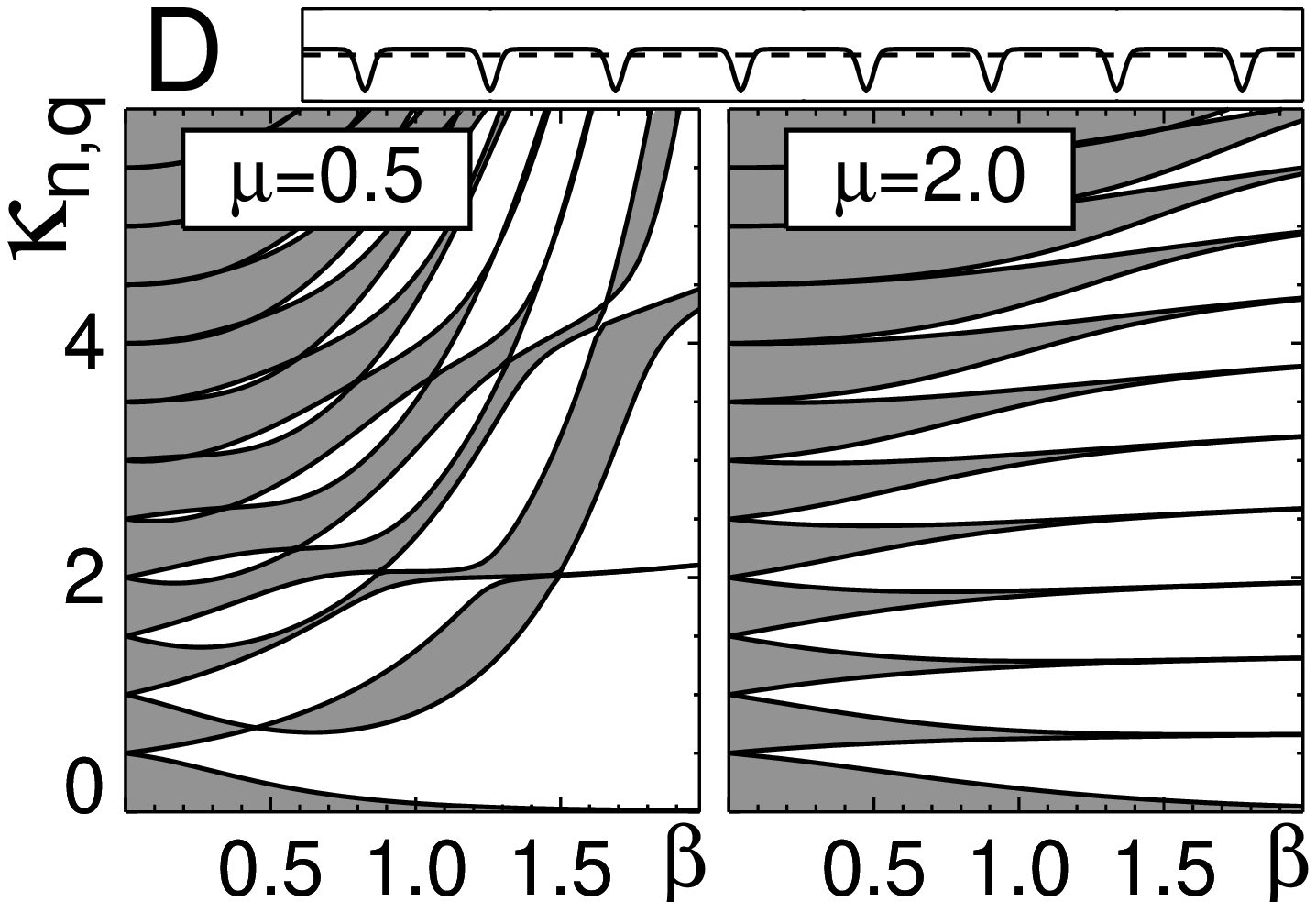}

\caption{\label{fig:fig2} Bandstructure $\kappa _{n,q}$ as a function of
$\beta $ for four different potentials. (A) a simple cosine potential
$V\propto \cos (x/\lambda )$, (B) the square wave potential, and
potentials given by $V\propto \pm (1+\cos (x/\lambda ))^{\gamma }$.
When $\gamma \gg 1$, the latter possess localized high potential
barriers (C) or localized potential wells (D). A Lévy flight ($\mu =1/2$)
is compared to ordinary diffusion ($\mu =2$). }
\end{figure}
In Eq.~(\ref{eq:bands}) $E_{n,q}$ are the eigenvalue bands labeled
by the discrete band index $n$ and the continuous Bloch phase $q$.
The spectrum of the system for vanishing potential is given by $E_{n,q}^{0}=|n/\lambda -q|^{\mu }$.
The Fourier coefficients of the periodic component of the eigenfunction
and the effective potential are given by $\fou{\theta }_{n}=1/2\pi \lambda \int _{2\pi \lambda }\dx \theta (x)\exp (-inx/\lambda )$
and $\fou{U}_{n}=1/2\pi \lambda \int _{2\pi \lambda }\dx U(x)\exp (-inx/\lambda )$,
respectively. The spectrum $E_{n,q}$ depends implicitly on $\beta $.
In the high temperature limit the eigenvalue bands merge to form a
continuous spectrum. For non-vanishing $\beta $ gaps between bands
emerge. The band structure determines the relaxation properties of
the system. For a comparison of different Lévy indices $\mu $, it
is more appropriate to compare the generalized crystal momentum defined
as $\kappa _{n,q}=E_{n,q}^{1/\mu }$.

Fig.~\ref{fig:fig2} depicts the $\beta $-dependence of the first
few bands $\kappa _{n,q}$ for different potentials, each one reflecting
a frequently encountered physical situation. In each panel ordinary
diffusion is compared to enhanced diffusion with Lévy index $\mu =1/2$.
The band structure of ordinary diffusion ($\mu =2$) in the cosine
potential displays only one significant gap~\cite{geise_00201:1979}
contrasting the superdiffusive case in which the narrowing effect
of individual bands is substantial (Fig.~\ref{fig:fig2}A). The effect
is even more pronounced in the square wave potential (Fig.~\ref{fig:fig2}B).
Band coupling leads to far more complex band structures when $\mu =1/2$.
The most striking difference occurs in the localized barrier (well)
potentials, Fig.~\ref{fig:fig2}C(D). In the example of copolymers
discussed above, these cases describe situations in which the polymer
consists mainly of a single type of monomer interspersed with small
intervals of another type of monomer at a higher or lower potential,
respectively. On one hand, the band structures are identical when
$\mu =2$, indicating that an ordinary diffusion process does not
distinguish between barriers and wells. On the other hand, if $\mu =1/2$,
the band structures differ considerably, the shape of the potential
has a profound impact on the band structure and thus on the dynamics
of the system. In a system containing barriers, a particle located
at any given position may initiate a distant jump with a probability
which is decreased by the concentration of energetically unfavorable
target positions. In the repetitive barrier potential this is low,
so the process is not considerably affected. In~Fig.~\ref{fig:fig2}D
the particle is likely to be trapped in a potential well. The probability
of initiating a distant transition to another energetically favorable
state is low and the repetitive well potential slows down the dispersion
dramatically. 

\begin{figure}
\noindent \includegraphics[  width=0.50\columnwidth]{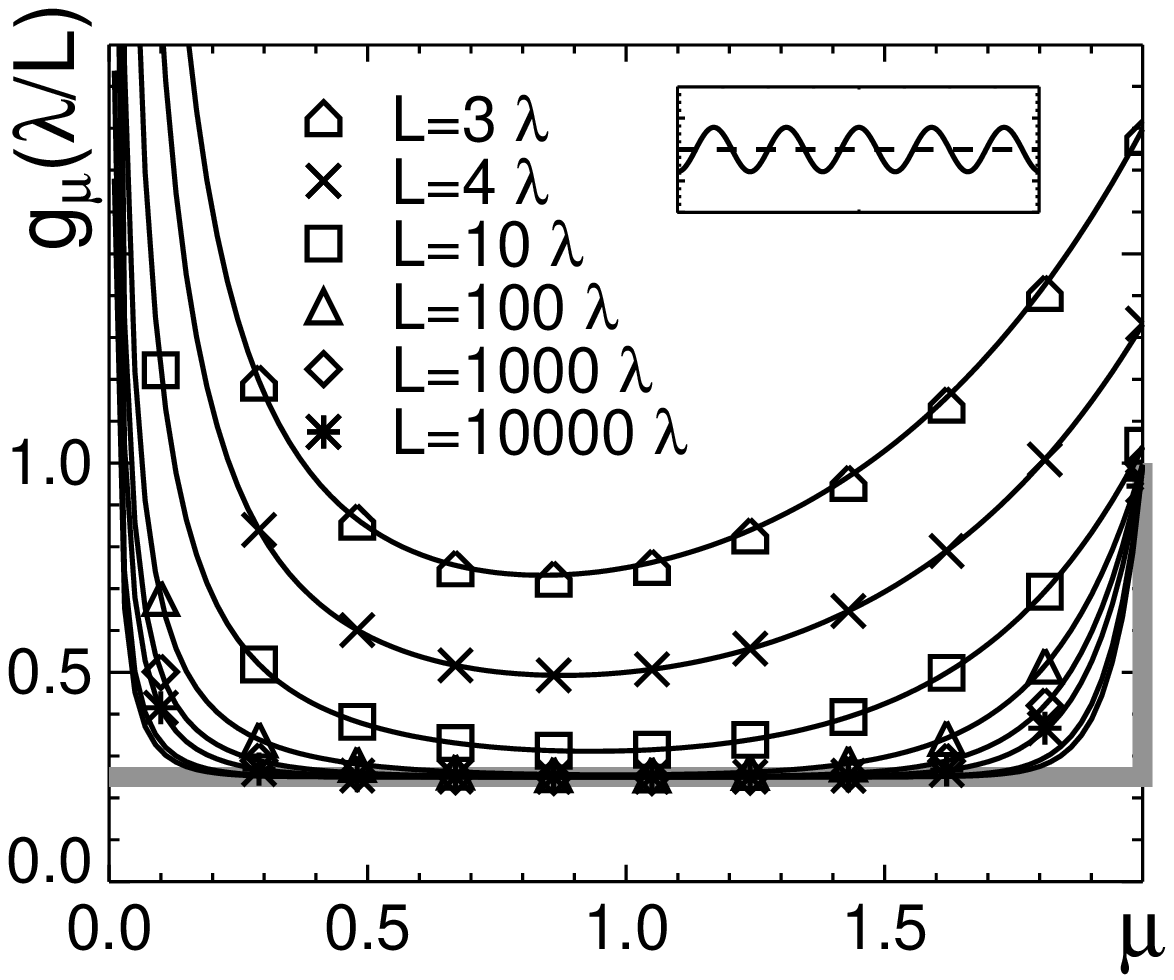}\hfill{}\includegraphics[  width=0.50\columnwidth]{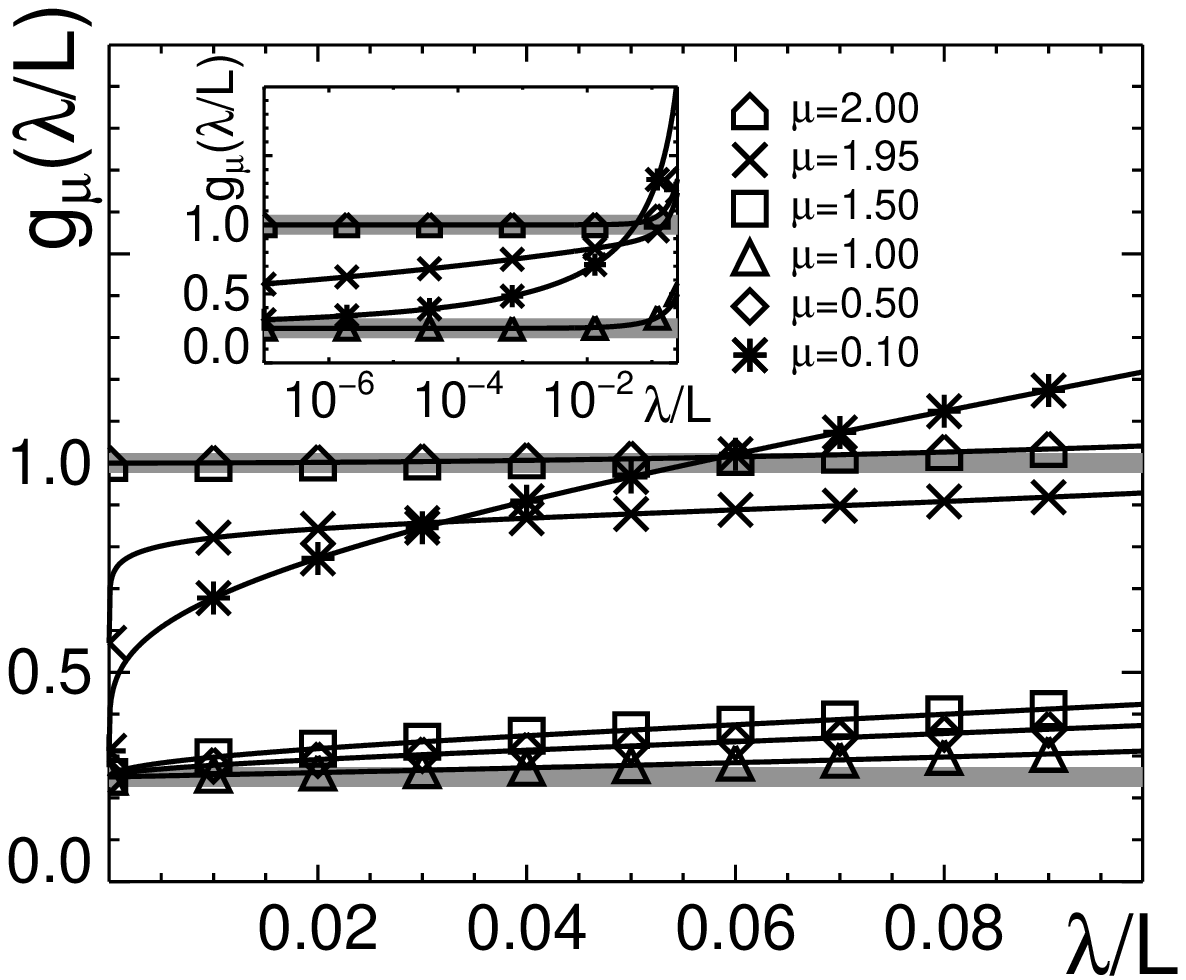}

\caption{\label{cap:fig3}The quantity $g_{\mu }(\lambda /L)$ in the high
temperature regime as a function of Lévy index $\mu $ and fixed system
size (left). The solid lines (symbols) depict the results obtained
from perturbation theory (numerics). Thick gray lines indicate the
asymptotic limit of $1/4$ ($1$) if $\mu <2$ ($\mu =2$). Viewed
as a function of inverse relative system size $\lambda /L$ for a
set of values of $\mu $ (right) indicates that convergence to the
limiting values is slowest when $\mu $ is small or slightly less
than $2$.}
\end{figure}
The asymptotic behavior is governed by the lowest band of the spectrum,
$n=0$, and $q\ll \lambda ^{-1}$. For $\beta =0$ the system is freely
superdiffusive. This yields $E_{0,q}^{0}=q^{\mu }$. When $\beta $
is finite the $q^{\mu }$-dependence remains, i.e. $E_{0,q}\approx D(\beta )\, q^{\mu }$.
The generalized diffusion coefficient $D(\beta )$, however, is reduced
to a value less than unity; the process is slowed down when a potential
is present. The high temperature regime can be investigated by expanding
the effective potential $U$ in~(\ref{eq:statschroedinger}) in powers
of $\beta $. Neglecting all terms of order higher than $\mathcal{O}(\beta ^{3})$
we obtain\begin{equation}
\fou{U}_{n}=\frac{1}{\lambda ^{\mu }}\left[\frac{\beta }{2}|n|^{\mu }\fou{V}_{n}-\frac{\beta ^{2}}{8}\sum _{m}\fou{V}_{n-m}\left(|n|^{\mu }-2|m|^{\mu }\right)\fou{V}_{m}\right].\label{eq:uofbeta}\end{equation}
Inserted into~(\ref{eq:bands}) yields in perturbation theory $E_{0,q}\approx q^{\mu }[1-\beta ^{2}G_{\mu }(q)]$.
As expected, the generalized diffusion coefficient $D(\beta )=1-\beta ^{2}G_{\mu }(q)$
decreases quadratically with increasing $\beta $. The factor $G_{\mu }(q)$
is positive and depends on $V$ and $\mu $. It quantifies the effect
on the asymptotics, the larger $G_{\mu }(q)$ the stronger the slowing
down effect of the potential. We obtain\begin{align}
G_{\mu }(q) & =\frac{1}{4}\sum _{m>0}|\fou{V}_{m}|^{2}g_{\mu }(q\lambda /m)\quad \text {where}\label{eq:grossgmu}\\
g_{\mu }(z) & =\frac{1}{z^{\mu }}\left(\frac{1}{(1-z)^{\mu }-z^{\mu }}+\frac{1}{(1+z)^{\mu }-z^{\mu }}-2\right).\label{eq:kleingmu}
\end{align}
Noting that $\sum _{m}|\fou{V}_{m}|^{2}=1$ the asymptotic limit is\begin{equation}
\lim _{q\rightarrow 0}G_{\mu }(q)=\begin{cases}
 1 & \quad \mu =2\\
 1/4 & \quad 0<\mu <2.\end{cases}\label{eq:asympg}\end{equation}
The asymptotics are the same for any type of potential. In addition,
the rhs of~(\ref{eq:asympg}) is independent of $\mu $ with a discontinuity
to a higher value on the margin $\mu =2$. 

The limit $q\rightarrow 0$ represents an idealized system of infinite
extent. In a finite finite system of size $2\pi L$, the Bloch phase
acquires discrete values $q=n/L$ with $n\in \mathbb{N}$. The relaxation
time is defined by inverse of the lowest eigenvalue, obtained by~(\ref{eq:grossgmu})
at $q=L^{-1}\ll \lambda ^{-1}$. The result is shown in Fig.~\ref{cap:fig3}
for the cosine potential. In this case, eq.~(\ref{eq:grossgmu})
implies $G_{\mu }(1/L)=g_{\mu }(\lambda /L)$. On the left, $g_{\mu }$
is depicted as a function of $\mu $ for a number of system sizes.
Surprisingly, the asymptotic limit is not attained uniformly on the
$\mu $-interval $(0,2]$. Even for very large systems $g_{\mu }$
exhibits a minimum at an intermediate value $\mu \approx 1$. Interestingly,
as $\mu \rightarrow 0$ the factor $g_{\mu }$ diverges. Although
small values of $\mu $ are equivalent to heavy tails in the transition
probability, the potential strongly influences the dynamics in that
range. We conclude that Lévy flights with intermediate values of $\mu $
are most robust when perturbed by an external field. This may explain
why Lévy flights with $\mu \approx 1$ are the most efficient when
employed in random search~\cite{viswa_00911:1999}. 

\begin{figure}[ht]
\noindent \includegraphics[  width=1.0\columnwidth]{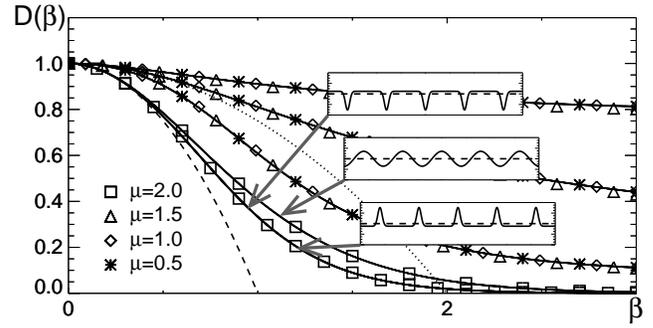}

\caption{\label{cap:fig4}The generalized diffusion coefficient $D(\beta )$
as a function of inverse temperature for a chosen set of Lévy indices
(indicated by the symbols in the lower left) and the set of potentials
depicted in the upper right and distinguished by color. The data were
obtained by numerical diagonalization of~\ref{eq:bands}. The dotted
($\mu <2$) and dashed ($\mu =2$) curves indicate the result obtained
by the perturbation expansion for small $\beta $.}
\end{figure}
Finally, we investigate the effective generalized diffusion coefficient
$D(\beta )$ in the low temperature regime. Since perturbation theory
fails here, we must rely on the numerical diagonalization of~(\ref{eq:bands}).
The result is depicted in figure~\ref{cap:fig4}. $D(\beta )$ is
identical for all superdiffusive processes in a given potential. However,
a comparison between potentials reveals a unique response of Lévy
flights to each potential shown in the inset. The effect on $D(\beta )$
is least pronounced in the potential barrier system, intermediate
for the cosine, and strongest in the potential well. In contrast,
ordinary diffusion shows a decrease in $D(\beta )$ which is not only
greater compared to all the other cases, but is independent of the
shape of the potential. For small $\beta $ the results are consistent
with those obtained from perturbation theory as indicated by the dotted
($\mu <2$) and dashed ($\mu =2$) lines.

In this Letter we have shown that Lévy flights are substantially affected
by external inhomogeneities, a feature generic to topologically induced
superdiffusion, absent in popular generalized Langevin models, yet
crucial for the understanding of physical applications such as protein
search on DNA strands and random motion on complex networks.

\begin{acknowledgments}
D. Brockmann thanks I.M. Sokolov and W. Noyes for interesting comments
and discussion.
\end{acknowledgments}
\bibliographystyle{apsrev}
\bibliography{/home/zwerg/bibliography/bib/books,/home/zwerg/bibliography/bib/paper}

\end{document}